\def\mev{{\rm MeV}}
\def\gev{{\rm GeV}}

\def\fm{{\rm fm}}

\def\qbar{\overline{q}}

\def\csw{c_{\rm sw}}

\def\za{Z_{\rm A}}
\def\zv{Z_{\rm V}}

\def\mq{m_{\rm q}}
\def\mP{m_{\rm P}}

\def\fb{f_B}
\def\fd{f_D}
\def\fbs{f_{B_s}}
\def\fds{f_{D_s}}
\def\frootm{f_P\sqrt{M_P}}
\def\bb{B_B}
\def\bbs{B_{B_s}}
\def\rgbb{\widehat{B}_B}
\def\rgbblo{\widehat{B}_B^{\rm LO}}
\def\rgbbnlo{\widehat{B}_B^{\rm NLO}}
\def\frootb{f_B\sqrt{\rgbb}}

\newcommand{\be}{\begin{equation}}
\newcommand{\ee}{\end{equation}}
\newcommand{\bea}{\begin{eqnarray}}
\newcommand{\eea}{\end{eqnarray}}
\newcommand{\eq}[1]{eq.\,(\ref{#1})}
\newcommand{\fig}[1]{Fig.\,\ref{#1}}
\newcommand{\tab}[1]{Table\,\ref{#1}}
\newcommand{\er}[2]{\,{}^{+#1}_{-#2}}

\input epsf
\input axodraw.sty
      [1997/09/01 v1.2b Il Nuovo Cimento]
\documentclass[varenna,cite]{cimento}
\begin{document}

 
\begin{flushright}
{\large Oxford Preprint: OUTP--97--59--P}\\
{\large October~1997}
\end{flushright}
 
\vspace*{15mm}
 
\begin{center}
{\huge Heavy quarks on the lattice: status and perspectives}\\[12mm]
 
{\large\bf Hartmut~Wittig\thanks{PPARC Advanced Fellow}}\\ [0.3cm]
Theoretical Physics, Oxford University,\\
1~Keble Road, Oxford OX1~3NP, UK
 
\end{center}
\vspace{15mm}
\begin{abstract}
    The lattice method for the calculation of weak decay amplitudes of
    heavy quark systems is introduced. Results for leptonic and
    semi-leptonic decays of heavy mesons and $B-\overline{B}$ mixing
    are reviewed.
\end{abstract}

\vspace{2cm}
 
\begin{center}
{\large To appear in Procs. of International School of Physics,
``Enrico Fermi''}\\[0.2 em]
{\large Course CXXXVII}\\[0.2 em]
{\large Heavy Flavour Physics: a probe of Nature's grand design}
\end{center}
 
\clearpage
 
 
\clearpage
\mbox{}
\vfill
\clearpage

\addtocounter{page}{-2}

\title{Heavy quarks on the lattice: status and perspectives}
\author{Hartmut Wittig}
\institute{Theoretical Physics, Oxford University,
                               1~Keble Road, Oxford OX1~3NP, UK
}

\maketitle

\section{Introduction: Heavy quarks on the lattice}

The violation of CP symmetry is one of the most important but still
least understood features of the Standard Model. Much effort has been
invested in order to analyse CP violation within the framework of the
Cabibbo-Kobayashi-Maskawa (CKM) pattern of quark mixing. Heavy quark
systems play an important r\^ole in the study of the CKM mixing
matrix: they contain information on its least known elements and serve
to test its parametrisation as a unitary $3\times3$ matrix in terms of
three angles and one CP violating complex phase.

The phenomenological extraction of CKM matrix elements involving heavy
quarks has been hampered by large hadronic uncertainties in the
evaluation of current matrix elements appearing in the relevant weak
decay amplitudes. Since quarks are confined within hadrons the
exchange of soft gluons makes a perturbative analysis of weak decays
impossible. During the past decade one has therefore sought to compute
these matrix elements non-perturbatively in lattice QCD.

Lattice simulations of QCD are by now a mature field. They allow for
the evaluation of hadronic masses, decay constants and form factors
from first principles. However, before lattice results can be applied
phenomenologically a critical assessment of systematic errors is
required. In this lecture I shall present lattice results for leptonic
and semi-leptonic decays of heavy quark systems as well as
$B-\overline{B}$ mixing. Systematic effects and methods how to
increase the accuracy of lattice data will be discussed. More detailed
information can be found in recent review
articles\,\cite{cts_jflynn_97,onogi_lat97,cb_sb97,wsc97}.

In order to formulate QCD on a discrete grid of points one
approximates space-time by a euclidean, hypercubic lattice with
lattice spacing~$a$ and volume $L^3\cdot T$. One then chooses a
discretisation of the QCD action involving the quark fields $q(x),
\qbar(x)$ and gauge fields $U_\mu(x)\in\rm SU(3),
\mu=1,\ldots,4$. One such discretisation was formulated by
Wilson\,\cite{wilson74} (``Wilson fermions'') and has been used for
almost all results I shall describe. Using the discretised QCD action
$S_{\rm QCD}$, one can define a partition function
\bea
  Z & = & \int D[U]D[\qbar]D[q]\,e^{-S_{\rm
  QCD}[U,\qbar,q]} \nonumber \\
    & = & \int D[U]\det Q\, e^{-S_G[U]},\label{EQ_partfunc}
\eea
where in the last line we have integrated out the quark fields,
resulting in the determinant of the Wilson-Dirac operator times the
exponentiated pure gauge action. The expectation value of an
observable $\cal O$ is defined as
\be
  \langle{\cal O}\rangle = Z^{-1}\int D[U]{\cal O}\det Q\, e^{-S_G[U]}. 
\ee
In a numerical simulation one evaluates $\langle{\cal O}\rangle$ by
generating a representative sample of $N_c$ gauge configurations in a
Monte Carlo procedure. The expectation value $\langle{\cal O}\rangle$
is then approximated by the sample average $\overline{\cal O}$
\be
  \langle{\cal O}\rangle \simeq \overline{\cal O} =
  \frac{1}{N_c}\sum_{i=1}^{N_c} {\cal O}_i,
\ee
where ${\cal O}_i$ is the value of the observable computed on the
$i$th configuration. Since one is working with a finite number of
configurations, $\overline{\cal O}$ is obtained with a statistical
error proportional to $1/\sqrt{N_c}$.  This procedure yields the
observable for non-zero values of the lattice spacing~$a$. The
continuum result is obtained in the limit $a\rightarrow0$. In
practice, this usually means that the simulation must be repeated for
several values of~$a$ so that the results can be extrapolated to
$a=0$.

The appearance of the determinant in \eq{EQ_partfunc}\ presents a
major obstacle for numerical simulations since it is highly non-local
and thus its evaluation is a huge computational overhead. Before the
advent of efficient algorithms it had been suggested to disregard its
effects completely by setting $\det Q=1$. This defines the so-called
{\it Quenched Approximation\/}\cite{Valence}, which in physical terms
corresponds to neglecting quark loops in the determination of
$\langle{\cal O}\rangle$. Although of unknown quality, the quenched
approximation is still widely used and also accounts for most of the
material covered here.

Since the (inverse) lattice spacing acts as an ultraviolet cutoff, its
value in physical units places constraints on the scales that one is
able to study. On present computers typical values lie in the range of
$a^{-1}=2-4\,\gev$. These relatively low values of the cutoff imply
that one expects large cutoff effects for the charm quark whose mass is
not too far below $a^{-1}\,[\gev]$. More importantly, the $b$ quark
cannot be studied directly since its mass lies above the
cutoff. The following methods are used used to circumvent this
problem:
\begin{itemize}
\itemsep 0pt
\item Reduction of lattice artefacts
\item Static approximation
\item Non-relativistic QCD (NRQCD)
\end{itemize}
In the first approach one seeks to cancel the leading cutoff effects
of order~$a$ in the Wilson action by employing so-called improved
actions and
operators\,\cite{SymanzikII,SW85,Heatlie91,paperI,paperIII}. For
instance, the O($a$) improved lattice action is defined as
\be
  S_{\rm QCD}^I[U,\qbar,q] =   S_{\rm QCD}[U,\qbar,q]
  + \csw\frac{ia}{4}\sum_{x,\mu,\nu}\qbar(x)\sigma_{\mu\nu}
			F_{\mu\nu}(x)q(x),
\ee
where $\csw$ is an improvement coefficient and $F_{\mu\nu}$ is a
lattice transcription of the field tensor. Further coefficients appear
in the definitions of the improved axial and vector currents. Provided
that all improvement coefficients are chosen appropriately one can
show that lattice artefacts of O($a$) are cancelled completely in
masses and matrix elements. In a different approach, which can be
applied for improved and unimproved actions, a reduction of lattice
artefacts is achieved through absorbing higher-order effects of the
quark mass into a rescaling factor\,\cite{EKM96}
\be
   q(x) \longrightarrow e^{a\mP/2}q(x),\qquad a\mP=\ln(1+a\mq),
\ee
where $a\mq$ is the bare quark mass. This normalisation is called the
El-Khadra-Kronfeld-Mackenzie (EKM) norm. It must be emphasised that
neither improvement nor the EKM norm solve the problem that the $b$
quark cannot be studied directly. However, they reduce lattice
artefacts around the charm quark mass, so that the obtained results
can be extrapolated to the $b$ quark mass much more reliably.

The static approximation is based on the leading term of the expansion
of the heavy quark propagator in powers of the inverse heavy quark
mass $1/m_Q$. Thus, one regards the $b$ quark as infinitely heavy in
this approach and expects that results will be subject to corrections
of order $\Lambda_{\rm QCD}/m_Q$. Finally, NRQCD is an effective
theory based on an expansion of the QCD action in the four-velocity of
the heavy (non-relativistic) quark. Again one expects corrections to
the effective theory whose influence on physics results has to be
assessed.

From this discussion it is clear that none of the above methods gives
an entirely satisfactory description of heavy quarks on the
lattice. However, they all provide complementary information which can
be used to reveal the full picture.

Besides lattice artefacts, another important source of systematic
errors is the explicit breaking of chiral symmetry by the Wilson
action. As a consequence, lattice versions of the local vector and
axial currents are not conserved. Instead they are related to their
continuum counterparts by finite renormalisations $\zv$ and $\za$,
respectively. Although $\zv$ and $\za$ have been computed
non-perturbatively for O($a$) improved
actions\,\cite{UKQCD_za,paperIV} these factors are known only in
one-loop perturbation theory in the unimproved case. Furthermore,
explicit chiral symmetry breaking causes operators with definite
chirality -- such as the four-fermion operator used to describe
$B-\overline{B}$ mixing -- to mix with operators of opposite
chirality. Therefore, several matrix elements have usually to be
determined on the lattice and subsequently matched to the continuum
matrix element. The general expression in the unimproved theory thus
is
\be
   \langle f|{\cal O}|i\rangle^{\rm cont}
  = \sum_{\alpha}\,Z_\alpha\langle f|{\cal O}_\alpha^{\rm latt}|i\rangle
  + O(a),
\ee
where the $Z_\alpha$'s are the appropriate normalisation factors, and
lattice artefacts of order~$a$ arise through mixing with higher
dimension operators.

Finally, lattice estimates of dimensionful quantities are subject to
uncertainties in the lattice scale. They are due to the fact that
different quantities like $f_\pi, M_\rho,\ldots$, which are commonly
used to set the scale $a^{-1}$ in physical units give different
results. This is closely related to using the quenched approximation,
since loop effects are not expected to be the same for different
quantities.

\section{Leptonic decays of heavy-light mesons}

\begin{figure}[tbp]
\vspace{-3.8cm}
\begin{center}
\epsfxsize=13 true cm
\epsfbox{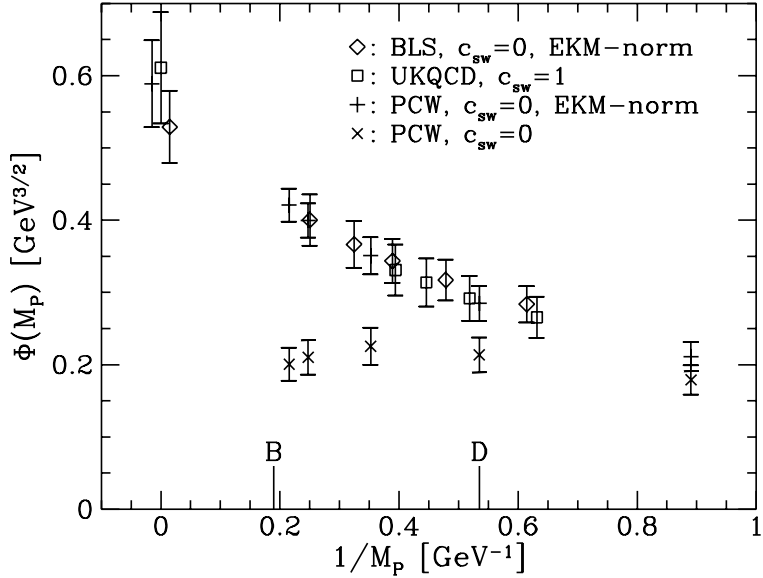}
\end{center}
\vspace{-2.0cm}
\caption{$\Phi(M_P)$ versus $1/M_P$ using data from different
collaborations for $a^{-1}\simeq3\,\gev$. The data from
\protect\cite{PCW_prop_93} are shown with (plus signs) and without
(crosses) including the EKM factors.}
\label{FIG_frootm}
\end{figure}

The leptonic decay constant $f_P$ of a heavy-light pseudoscalar meson
is related to the matrix elements of the axial current on the lattice
via
\be
   \langle0|A_4(0)|{\rm PS}\rangle = M_Pf_P/\za,
\ee
where $M_P$ is the pseudoscalar mass and $\za$ is the renormalisation
factor of the lattice axial current. Both the matrix element and $M_P$
are obtained from the asymptotic behaviour of the euclidean
correlation function of the axial current at large separation~$t$
\be
  \sum_{\vec{x}}\langle A_4(\vec{x},t)A_4^\dagger(0)\rangle 
  \stackrel{t\gg0}{\simeq}\frac{|\langle0|A_4(0)|{\rm PS}
  \rangle|^2}{2M_P}\left\{e^{-M_Pt} + e^{-M_P(T-t)}\right\}.
\ee
The decay constant $f_P$ can then be studied at several different
values of the mass of the heavy quark. This enables one to study some
predictions by the Heavy Quark Effective Theory (HQET). It is well
known that HQET predicts the following scaling law in the limit of an
infinitely heavy quark
\be
   \frootm\stackrel{m_Q\to\infty}{\longrightarrow}
   {\rm const}\times\alpha_s(M_P)^{-2/\beta_0},
\label{EQ_frootm}
\ee
where $\alpha_s$ is the strong coupling constant and
$\beta_0=11-2n_f/3$.  In order to test the quality of this prediction
we plot in \fig{FIG_frootm} the quantity
\be
   \Phi(M_P) = f_P\sqrt{M_P}\left(
   \frac{\alpha_s(M_P)}{\alpha_s(M_B)}\right)^{2/\beta_0}
\ee
as a function of $1/M_P$, using data in the static
approximation\,\cite{PCW_stat_92,BLS93,bbar} and for relativistic
heavy quarks\,\cite{BLS93,quenched,PCW_prop_93}. Using the static
approximation as the limiting case, the figure illustrates that there
are large corrections in $1/M_P$ to the scaling law, provided that
lattice artefacts have been treated, either by using improvement
(i.e. $\csw\ge1$) or by employing the EKM norm. Failure to address the
problem of lattice artefacts leads to an inconsistent mass behaviour
of the data in the static and relativistic regimes. (c.f. crosses in
\fig{FIG_frootm}). We conclude that the HQET scaling law is not
satisfied at the physical $B$ and $D$ meson masses and that the
treatment of lattice effects is crucial for the computation of
heavy-light decay constants in general. Moreover, this example
illustrates the interplay between different formulations of heavy
quarks.

\begin{table}
\begin{center}
\begin{tabular}{lcllll}
\hline
Collab. & $a\,[\fm]$ & $\fd\,[\mev]$ & $\fds\,[\mev]$ & $\fb\,[\mev]$
                     & $\fbs\,[\mev]$ \\
\hline
FNAL$^\ast$ \cite{FNAL_lat97} & 0 & 205(9)(27) & 215(7)(30) & 166(10)(28)
                & 1.17(4)(3)  \\
JLQCD$^\ast$ \cite{JLQCD_lat97} & 0 & 192(10)$\er{11}{16}$ &
           213(11)$\er{12}{18}$ & 163(12)$\er{13}{16}$ & \\[0.2 em]
MILC$^\ast$ \cite{MILC_lat97} & 0 & 186(10)$\er{27}{18}\er{9}{0}$ &
           199(8)$\er{40}{11}\er{10}{~0}$ &
           153(10)$\er{36}{13}\er{13}{~0}$ &
           1.10(2)$\er{5}{3}\er{3}{2}$\\ 
APE \cite{APE_fB97} & 0.07 & 221(17) & 237(16) & 180(32) & 1.14(8) \\
LANL \cite{LANL_fD} & 0.09 & 229(7)$\er{20}{16}$ & 260(4)$\er{27}{22}$
                    &  &  \\
PCW \cite{PCW_prop_93} & 0 & 170(30) &  & 180(50) & 1.09(2)(5) \\
UKQCD \cite{quenched} & 0.07 & 185$\er{4}{3}\er{42}{~7}$ &
                     212(4)$\er{46}{~7}$ & 160(6)$\er{59}{19}$ &
                     1.22$\er{4}{3}$  \\
BLS \cite{BLS93} & 0.06 & 208(9)(35)(12) &  230(5)(10)(19) &
                  187(10)(34)(15) & 1.11(6) \\
\hline
Hirosh.$^\dagger$ \cite{Hirosh97} & 0.12 & & & 184(7)(5)(37)(37) &
                  1.23(3)(3) \\ 
GLOK$^\dagger$ \cite{GLOK_qu_97} & 0.09 & & & 183(32)(28)(16) & 1.17(7) \\
SGO$^\dagger$ \cite{SGO_dyn_97} & 0.10 & & & 126--166 & 1.24(4)(4) \\
\hline
\end{tabular}
\end{center}
\caption{Results for heavy-light decay constants from different
collaborations. Data marked by an asterisk are preliminary. Results
obtained using NRQCD are marked by a dagger. All other collaborations
have used relativistic heavy quarks. The convention $f_\pi=131\,\mev$
is understood.}
\label{TAB_fP}
\end{table}

Recent results for heavy-light decay constants using relativistic
heavy quarks and NRQCD are compiled in \tab{TAB_fP}. Besides the
statistical error, most groups now quote one or more systematic
errors.  Note, however, that the estimation of systematic errors can
vary significantly among different collaborations. The results shown
in \tab{TAB_fP} are broadly consistent, but the errors are still large
and dominated by systematic effects. Despite the apparent consistency
of the data it would be premature to simply combine them into global
estimates of decay constants, because the details of the analysis of
the raw lattice data can differ substantially among different
groups. A common analysis of lattice data from many collaborations has
been described in\,\cite{wsc97}. There the aim was to perform the
extrapolation to the continuum limit and to present a uniform
estimation of systematic errors from a number of sources. These
include the choice of lattice scale (e.g. $f_\pi, M_\rho,\ldots$), the
quark field normalisation, uncertainties in the perturbative values of
$\za$, and variations in fitting and extrapolation procedures. The
results in the continuum limit can be summarised as
\bea
  \fd & = & 191\pm19\,({\rm stat})\er{~3}{20}\,({\rm syst})\,\mev,
\qquad \fds = 206\pm17\,({\rm stat})\er{~6}{22}\,({\rm syst})\,\mev \\
  \fb & = & 172\pm24\,({\rm stat})\er{13}{19}\,({\rm syst})\,\mev,
\qquad \fbs/\fb = 1.14(8)
\label{EQ_fB}
\eea
Further details and comparisons to other theoretical results can be
found in\,\cite{wsc97}.

\section{$B-\overline{B}$ mixing}

\begin{figure}[h]
\begin{center}
\vspace{-1.5cm}
\begin{picture}(350,200)(0,0)
\ArrowLine(0,130)(30,130)
\ArrowLine(30,130)(90,130)
\ArrowLine(90,130)(120,130)
\ArrowLine(30,70)(0,70)
\ArrowLine(90,70)(30,70)
\ArrowLine(120,70)(90,70)
\DashLine(30,130)(30,70){4}
\DashLine(90,130)(90,70){4}
\put(128,128){\makebox(10,10)[l]{$d$}}
\put(128,68){\makebox(10,10)[l]{$\overline{b}$}}
\put(-12,128){\makebox(10,10)[r]{$b$}}
\put(-12,68){\makebox(10,10)[r]{$\overline{d}$}}
\put(42,133){\makebox(30,10)[]{{$u,c,{t}$}}}
\put(43,60){\makebox(30,10)[]{{$u,c,{t}$}}}
\put(18,95){\makebox(10,10)[r]{{$W$}}}
\put(94,95){\makebox(10,10)[l]{{$W$}}}
\ArrowLine(220,130)(250,130)
\DashLine(250,130)(310,130){4}
\ArrowLine(310,130)(340,130)
\ArrowLine(250,70)(220,70)
\DashLine(340,70)(250,70){4}
\ArrowLine(340,70)(310,70)
\ArrowLine(250,130)(250,70)
\ArrowLine(310,70)(310,130)
\put(348,128){\makebox(10,10)[l]{$d$}}
\put(348,68){\makebox(10,10)[l]{$\overline{b}$}}
\put(208,128){\makebox(10,10)[r]{$b$}}
\put(208,68){\makebox(10,10)[r]{$\overline{d}$}}
\put(277,133){\makebox(10,10)[]{{$W$}}}
\put(215,95){\makebox(30,10)[r]{{$u,c,{t}$}}}
\put(277,60){\makebox(10,10)[]{{$W$}}}
\put(315,95){\makebox(30,10)[l]{{$u,c,{t}$}}}
\end{picture}
\end{center}
\vspace{-2.5cm}
\caption{Box diagram contributions to $B^0-\overline{B}^0$ mixing.}
\label{FIG_box}
\end{figure}
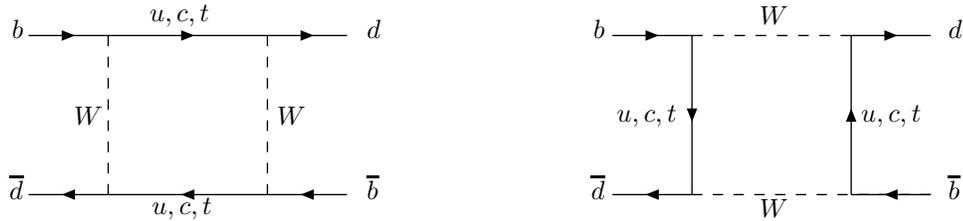

We now turn our attention to matrix elements which are used to
describe oscillations between $B^0$ and $\overline{B}^0$ states. The
mechanism of $B^0-\overline{B}^0$ mixing is illustrated by the box
diagrams shown in \fig{FIG_box}.

The mass difference $\Delta m$ between the mass eigenstates $B^0$ and
$\overline{B}^0$ is related to the CKM matrix elements $V_{td}$ and
$V_{tb}$ via
\be
  \Delta m\propto |V_{td}V_{tb}^*|^2\,M_Bf_B^2\rgbb.
\label{EQ_Deltam}
\ee
Here, $\fb$ is the decay constant of the $B$ meson encountered in the
previous section, and $\bb$ denotes the $B$ parameter defined by
\be
   \bb(\mu) = 
\frac{\langle\overline{B}^0|\widehat{O}_L(\mu)|B^0\rangle}
     {\frac{8}{3}\fb^2M_B^2},
\ee
where $\mu$ is the renormalisation scale and the four-fermion operator
$\widehat{O}_L$ is given by
\be
   \widehat{O}_L = \left(\overline{b}\gamma_\mu(1-\gamma_5)d\right)\,
                   \left(\overline{b}\gamma_\mu(1-\gamma_5)d\right).
\ee
The ``hat'' on $\bb$ in \eq{EQ_Deltam} signifies that the dependence
of $\bb$ on the renormalisation scale has been divided out. The
resulting renormalisation group invariant $B$ parameter can be defined
at leading (LO) or next-to-leading order (NLO) in $\alpha_s$ via
\bea
  \rgbblo & = & \alpha_s(\mu)^{-2/\beta_0}\bb(\mu) \\
  \rgbbnlo& = &
  \alpha_s(\mu)^{-2/\beta_0}\left(1+\frac{\alpha_s(\mu)}{4\pi}J_5\right)
  \bb(\mu),
\eea
where $J_5$ is derived from the one- and two-loop anomalous dimensions
of the operator $\widehat{O}_L$. It is important to realise that the
combination $\fb^2\rgbb$ is the principal unknown quantity which
relates the experimentally measured mass difference $\Delta m$ to the
CKM matrix elements in \eq{EQ_Deltam}. Thus, $\fb^2\rgbb$ is an
important ingredient for the study of CP violation.

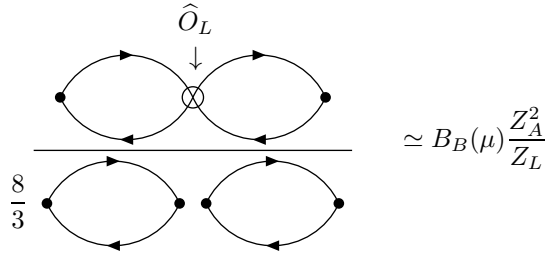
\begin{figure}[h]
\begin{center}
\vspace{-1.4cm}
\begin{picture}(350,200)(0,0)
\Vertex(60,120.00){2}
\Vertex(160,120.00){2}
\GCirc(110,120.00){4}{4}
\ArrowArcn(85,108.34)(27.59,155,25)
\ArrowArcn(85,131.66)(27.59,335,205)
\ArrowArcn(135,108.34)(27.59,155,25)
\ArrowArcn(135,131.66)(27.59,335,205)
\put(110,145){\makebox(3,10)[]{$\widehat{O}_L$}}
\put(110,130){\makebox(3,10)[]{$\downarrow$}}
\Line(50,100)(170,100)
\Vertex(55,80.00){2}
\Vertex(105,80.00){2}
\ArrowArcn(80,68.34)(27.59,155,25)
\ArrowArcn(80,91.66)(27.59,335,205)
\Vertex(115,80.00){2}
\Vertex(165,80.00){2}
\ArrowArcn(140,68.34)(27.59,155,25)
\ArrowArcn(140,91.66)(27.59,335,205)
\put(45,76){\makebox(3,10)[r]{$\displaystyle{{8}\over{3}}$}}
\put(190,100){\makebox(40,10)[l]{$\simeq
B_B(\mu)\displaystyle{{Z_A^2}\over{{Z_L}}}$}}
\end{picture}
\end{center}
\vspace{-2.4cm}
\caption{Lattice measurable for the contribution of $\widehat{O}_L$ to
the $B$ parameter (see text). Full dots represent insertions of the
axial current.} 
\label{FIG_BBlatt}
\end{figure}

On the lattice $\bb(\mu)$ is obtained from a ratio of two- and
three-point functions as depicted in \fig{FIG_BBlatt}.  Apart from the
renormalisation constant $\za$ of the lattice axial current the
relation between the lattice and continuum versions of $\bb$ will also
contain the factor $Z_L$ associated with the lattice version of
$\widehat{O}_L$. Furthermore, as there is mixing between
$\widehat{O}_L$ and other four-fermion operators with opposite and
mixed chiralities, all contributions have to be evaluated and matched
to the continuum theory. Since the relevant $Z$-factors have only been
computed in one-loop perturbation theory, the matching procedure
introduces considerable uncertainties into the final estimates of the
$B$ parameter. For lattice results obtained in the static
approximation and using $\csw=1$, these uncertainties have been
estimated to be as large as 25\%\,\cite{bbar,GMbbar,wsc97}.

\begin{table}
\begin{center}
\begin{tabular}{lclll}
\hline
Collab. & $a\,[\fm]$ & $\bb(m_b)$ & $\rgbb^{\rm LO}$ &
$\rgbb^{\rm NLO}$ \\
\hline
Kentucky$^\dagger$ \cite{Ken_bbar} & 0.09  & 0.97(4) & 1.42(6) & 1.54(6) \\
G+M$^\dagger$ \cite{GMbbar} & 0.09 & 0.63(4) & 0.92(6) & 1.00(6) \\
                            & 0.09 & 0.73(4) & 1.07(6) & 1.16(6)
                               \\
UKQCD$^\dagger$ \cite{bbar} & 0.07 & 0.69$\er{3}{4}\er{2}{1}$ &
1.02$\er{5}{6}\er{3}{2}$ & 1.10$\er{5}{6}\er{3}{2}$ \\[0.2 em]
UKQCD$^\dagger$ \cite{bbar_rev} & 0.07 & 0.83$\er{3}{4}\er{2}{1}$ &
		& 1.32$\er{5}{7}\er{3}{2}$ \\
\hline
JLQCD$^*$ \cite{JLQCD_bbar} & 0.06 & 0.840(60) & 1.23(9) & 1.34(10) \\
                            & 0.08 & 0.895(47) & 1.31(7) & 1.42(8)
                        \\
B+S$^*$ \cite{soni_lat95} & 0 & 0.89(6)(4) & 1.30(9)(6) & 1.42(10)(6) \\
ELC \cite{abada92} & 0.05 & 0.84(5) & 1.24(7) & 1.34(8) \\
BDHS \cite{BDHS88} & 0.08 & 0.93(14) & 1.36(20) & 1.48(22) \\
\hline
\end{tabular}
\end{center}
\caption{Data for the $B$ parameter from different collaborations at a
reference scale $m_b=5\,\gev$. Data marked by a dagger have been
obtained using the static approximation. Data marked by an asterisk
are preliminary.}
\label{TAB_BB}
\end{table}

Lattice data for $\bb(\mu)$ have been published for propagating and
static heavy quarks and are shown in \tab{TAB_BB}. It is clear that
lattice data for $\bb$ do not yet allow for a continuum extrapolation
as in the case of $\fb$. Instead one may quote a common estimate with
an error that encompasses the spread of different results. In
ref.\,\cite{wsc97} the result is
\be
   \bb(5\,\gev) = 0.85\er{13}{22},\qquad \rgbb=1.3\er{2}{3}.
\ee
The global result for the ratio $\bbs/\bb$ quoted in \cite{wsc97} is
\be
   \bbs/\bb = 1.00\pm0.02.
\ee
We can now combine the results for the decay constant $\fb$ in
\eq{EQ_fB} with that for the $B$ parameter. Combining the errors in
quadrature one finds
\be
   \frootb=195\er{30}{40}\,\mev.
\ee
For the SU(3)-flavour breaking ratio involving both the decay
constants and $B$ parameters the global result in \cite{wsc97} is 
\be
   \frac{\fbs^2\bbs}{\fb^2\bb}=1.38\pm0.15.
\ee
These estimates can now be used in the study of the CKM matrix and CP
violation. 

\section{Semi-leptonic $B\to\pi$ and $B\to\rho$ decays}

We will now discuss lattice results for semi-leptonic decays of $B$
mesons. Due to lack of space we will concentrate on heavy-to-light
transitions. Semi-leptonic $B\to D$ and $B\to D^*$ decays are
reviewed, for instance, in ref.\,\cite{cts_jflynn_97}.

Decays like $\overline{B}^0\to\pi^+\ell^-\overline{\nu}_\ell$ or
$\overline{B}^0\to\rho^+\ell^-\overline{\nu}_\ell$ have attracted much
interest recently, since they can be used to extract $V_{ub}$, which
is one of the most poorly known CKM matrix elements. Compared to the
pseudoscalar decay constant or the $B$ parameter, the study of
semi-leptonic decays is more complicated due to the kinematics
involved in their description. The relevant matrix elements of the
weak $V-A$ current are parametrised in terms of form factors which
depend on the momentum transfer $q^2$ between the initial and final
mesons. For instance, if the final state is a pseudoscalar meson only
the vector current contributes, and there are two independent form
factors $f^+$ and $f^0$
\be
  \left\langle {\rm PS}(k)\left|V_\mu\right|B(p)\right\rangle
  = f^+(q^2)\left\{(p+k)_\mu-\frac{M_B^2-M_P^2}{q^2}q_\mu\right\}
   +f^0(q^2)\frac{M_B^2-M_P^2}{q^2}q_\mu,
\ee
where $q_\mu=p_\mu-k_\mu$. If the final state is a vector meson, there
are four form factors $V(q^2), A_1(q^2), A_2(q^2)$ and $A_0(q^2)$,
associated with matrix elements of the vector and axial vector
currents, respectively. 

The need to control lattice artefacts places constraints on the
possible values of lattice momenta $\vec{p}$ and $\vec{k}$. Together
with the constraints on the heavy quark mass this implies that one
usually obtains form factors for typical momenta
$|\vec{p}\,|\leq1.5\,\gev/c$ and heavy quark masses in the region of
charm. The ``generic'' heavy-to-light semi-leptonic decay in lattice
simulations is thus $D\to K\ell\nu_\ell$ for momentum transfers in the
range $-1\,\gev^2/c^2\leq q^2\leq 2\,\gev^2/c^2$. Although in
principle one is interested in the $q^2$-dependence of form factors,
they are commonly quoted at $q^2=0$. Frequently a simple pole ansatz
is used to model the $q^2$-dependence:
\be
   F(q^2) = \frac{F(0)}{(1-q^2/M_{\rm pole}^2)^{n_F}}.
\label{EQ_pole}
\ee
Here, $F$ is a generic form factor, $M_{\rm pole}$ is of the order of
the heavy-light meson mass, and $n_F=0,1,2,\ldots$ parametrises
constant, monopole, dipole and higher multipole behaviour of $F$. A
recent compilation of lattice results for form factors for
semi-leptonic $D$ decays can be found in
\cite{cts_jflynn_97,jns_lat95}. Since the lattice form factors for
these decays are determined in a region around $q^2=0$, the pole
ansatz in \eq{EQ_pole} merely serves to guide the interpolation of
$F(q^2)$ to $q^2=0$, so that no model dependence is introduced.

The situation changes significantly if one considers semi-leptonic
$B\to\pi$ or $B\to\rho$ decays. Here the form factors are obtained
through extrapolation in the heavy quark mass to the mass of the $b$
quark. Since similarly large values of lattice momentum $|\vec{p}\,|$
cannot be considered due to restrictions imposed by lattice artefacts,
the accessible region of $q^2$ is pushed to large values near $q_{\rm
max}^2$. This in turn leaves a long and potentially uncontrollable
extrapolation in $q^2$ to determine $F(0)$. In order to map out the
$q^2$-behaviour one usually cannot avoid relying on model assumptions.

We shall now describe how the $q^2$-behaviour can be constrained by
requiring consistency with heavy quark symmtery (HQS), kinematical
constraints and scaling laws implied by light-cone sum rules.  In
analogy to the decay constant in \eq{EQ_frootm}, HQS predicts the
following leading scaling behaviour of form factors in the infinite
mass limit at fixed values of $\omega$ (which is the product of
four-velocities of the initial and final mesons):
\be
  f^+(\omega)\sim M^{ 1/2},\quad
  f^0(\omega)\sim M^{-1/2},\quad
  V(\omega) \sim  M^{ 1/2},\quad
  A_1(\omega)\sim M^{-1/2},\ldots
\ee
Here, $M$ is the mass of the heavy-light meson. The extrapolations
of form factors to the $b$ quark mass can be performed using model
functions motivated by the above scaling laws. Additional scaling laws
are provided by light-cone sum rule analyses
\cite{AliBrauSim_94,BallBrau_97}, which predict that all form factors
scale like $F\sim M^{-3/2}$ at $q^2=0$. In the heavy quark limit
\be
   1-\frac{q^2}{M_{\rm pole}^2} \sim \frac{1}{M},
\ee
and thus the scaling laws predicted by HQS and light-cone sum rules
can be combined to infer a value for $n_F$ in the pole formula
\eq{EQ_pole}. Also, kinematical constraints at $q^2=0$ such as
\be
  f^+(0)=f^0(0)
\label{EQ_kincon}
\ee
can be used to analyse the $q^2$-dependence of lattice form
factors. In \tab{TAB_FF} we list lattice estimates for form factors
for $\overline{B}^0\to\pi^+\ell^-\overline{\nu}_\ell$ and
$\overline{B}^0\to\rho^+\ell^-\overline{\nu}_\ell$ decays. It should
be noted that only the UKQCD data are consistent with all constraints
discussed above.

\begin{table}
\begin{center}
\begin{tabular}{lcllll}
\hline
Collab. & $a\,[\fm]$ & $f^+(0)$ & $V(0)$ & $A_1(0)$ & $A_2(0)$ \\
\hline
UKQCD \cite{FF_97} & 0.07 & 0.27(11) & $0.35\er{6}{5}$ & $0.27\er{5}{4}$
		   & $0.25\er{5}{3}$ \\
GSS \cite{GSS_FF}  & 0.06 & 0.43(19) & 0.65(15) & 0.28(3) & 0.46(23)
		   \\
APE \cite{APE_semil} & 0.09 & 0.35(8) & 0.53(31) & 0.24(12) & 0.27(80)
		   \\
ELC \cite{ELC_semil} & 0.05 & 0.30(14)(5) & 0.37(11) & 0.22(5) &
		   0.49(21)(5) \\
\hline
\end{tabular}
\end{center}
\caption{Lattice results for form factors for semi-leptonic $B\to\pi$
		   and $B\to\rho$ decays.}
\label{TAB_FF}
\end{table}

A very different approach to constrain the $q^2$-dependence in a
model-independent fashion has been discussed by Lellouch
\cite{lpl_B_to_pi}. Lattice data for the form factors $f^+$ and $f^0$
for $\overline{B}^0\to\pi^+\ell^-\overline{\nu}_\ell$ obtained near
$q^2_{\rm max}$ have been combined with dispersion relations and the
kinematical constraint \eq{EQ_kincon}. The method relies on
perturbative QCD in the evaluation of the dispersion relations and
general properties such as unitarity, analyticity and
crossing. However, existing lattice data for the form factors are at
present not precise enough in order to allow for stringent bounds at
$q^2=0$.

\begin{figure}[tbp]
\vspace{-3.8cm}
\begin{center}
\epsfxsize=13 true cm
\epsfbox{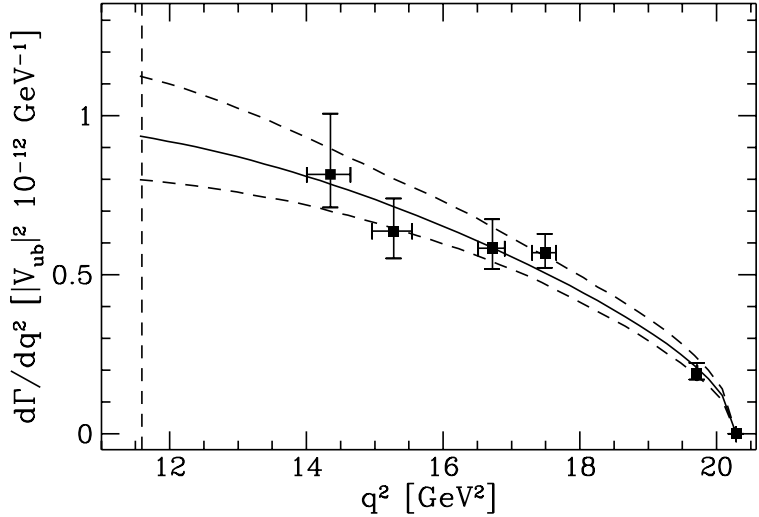}
\end{center}
\vspace{-2.8cm}
\caption{Differential decay rate for
$\overline{B}^0\to\rho^+\ell\overline{\nu}_\ell$ from
\protect\cite{B_to_rho}. Points are lattice data, and the fit and
variation in \protect\eq{EQ_rate} is represented by the solid and
dashed curves, respectively. The vertical dashed line marks the
endpoint of charm production.} 
\label{FIG_rate}
\end{figure}

Another proposal to avoid model dependence in the extraction of
$V_{ub}$ was made in \cite{B_to_rho}. Here one concentrates on the
{\it exclusive\/} decay
$\overline{B}^0\to\rho^+\ell^-\overline{\nu}_\ell$ in the region near
$q^2_{\rm max}$. Instead of attempting to extract the form factors at
$q^2=0$ one parametrises the differential decay rate by
\be
  \frac{d\Gamma}{dq^2}=10^{-12}
  \frac{G_F^2|V_{ub}|^2}{192\pi^3M_B^3}q^2\sqrt{\lambda(q^2)}
  {\cal A}^2(1+{\cal B}\big(q^2-q^2_{\rm max})\big),
\ee
where ${\cal A}$ and ${\cal B}$ are parameters and $\lambda$ is a
phase-space factor. The combination ${\cal A}^2(1+{\cal
B}\left(q^2-q^2_{\rm max})\right)$ parametrises the long-distance
hadronic dynamics, and ${\cal A}^2$ provides the overall
normalisation. Using lattice data for the form factors to evaluate the
differential decay rate the authors of \cite{B_to_rho} find
\be
  {\cal A} = 4.6\er{0.4}{0.3}\pm0.6\,\gev,\qquad
  {\cal B} = (-8\er{4}{6})\cdot10^{-2}\,\gev^2.
\label{EQ_rate}
\ee
The corresponding prediction of the decay rate is shown in
\fig{FIG_rate}. Given sufficient experimental data for the decay rate
in conjunction with accurate lattice results, a determination of
$V_{ub}$ will be possible.

\acknowledgments

I am grateful to Profs. I.~Bigi and L.~Moroni for the kind hospitality
at this school. I would like to thank J. Flynn, L. Lellouch,
J. Nieves, C. Sachrajda and other members of the UKQCD
Collaboration. The support of PPARC through the award of an Advanced
Fellowship is gratefully acknowledged.

\end{document}